\documentclass[12pt,a4paper]{article}
\usepackage{a4wide}
\usepackage{amsmath}
\usepackage{amssymb}
\usepackage{epsfig}
\usepackage{subfigure}
\usepackage{exscale}
\usepackage{float}
\usepackage{bbm}
\usepackage[numbers,sort&compress]{natbib}
\usepackage{pst-plot, pstricks,pst-math}
\usepackage{fancybox,amssymb,color}
\usepackage{graphicx}
\usepackage{pstricks, color, graphicx, epsfig, psfrag}
\usepackage{amsfonts,amsmath,amssymb,slashed}
\usepackage{dsfont}
\usepackage{bbm,bm}
\usepackage{fancyhdr, a4wide}
\usepackage[english]{babel}
\usepackage{subfigure}
\makeatletter
\@addtoreset{equation}{section}
\makeatother

\title{Diamagnetic and Paramagnetic Phases in Low-Energy Quantum Chromodynamics}

\author{Christoph P.\ Hofmann$^a$ \\ \\
\normalsize{$^a$ Facultad de Ciencias, Universidad de Colima} \\
\vspace{0.3cm}
\normalsize {Bernal D\'iaz del Castillo 340, Colima C.P.\ 28045, Mexico} \\}

\begin{document}

\maketitle

\begin{abstract} \normalsize

While it is known that the QCD vacuum in a magnetic background exhibits both diamagnetic and paramagnetic characteristics in the low-energy
domain, a systematic investigation of the corresponding phases emerging in the pion-dominated regime is still lacking. Here, within
two-flavor chiral perturbation theory, taking into account the pion-pion interaction, we analyze the subtle interplay between zero- and
finite-temperature portions in the magnetization and magnetic susceptibility. The dependence of the magnetic susceptibility on temperature
and magnetic field strength in the paramagnetic and diamagnetic phase is non-monotonic. Our low-energy analysis complements lattice QCD
that is currently operating at higher temperatures and stronger magnetic fields.

\end{abstract}

\maketitle

\section{Introduction}
\label{Intro}

Achieving a more quantitative understanding of the phases emerging in quantum chromodynamics (QCD) subjected to external magnetic fields,
is a major theme in current strong interaction research. One objective is to gain more rigorous insights into the nonperturbative regime of
QCD -- or the standard model of particle physics in general. Apart from theoretical aspects such as the characterization of compact neutron
stars or the evolution of the early universe, the problem is also phenomenologically relevant in view of heavy-ion collision experiments
that probe the quark-gluon plasma. In all these cases the magnetic fields are strong. Here we rather focus on the low-energy domain where
magnetic fields are weak and temperatures are low compared to the chiral symmetry breaking scale $\Lambda_{\chi} \approx 1 \, \text{GeV}$.

The thermomagnetic properties of QCD are described by the magnetization ${\mathfrak M}$ and the magnetic susceptibility $\chi$. The
classification of the QCD vacuum into diamagnetic or paramagnetic relies on the sign of the magnetic susceptibility, defined as the
response of the magnetization with respect to the external magnetic field $H$,
\begin{equation}
\chi = \frac{\mbox{d} {\mathfrak M}}{\mbox{d} |qH|} \, , \qquad {\mathfrak M} = - \frac{\mbox{d} z}{\mbox{d} |qH|} \, .
\end{equation}
Here $z$ is the free energy density and $q$ is the electric charge. Aside from lattice QCD simulations
\citep{BBCCEKPS12c,BEMNS13,BBEGS13,BBES13,BEMNS13b,BBEKS14,BEMNS14,BBES14,LT14,End14,BEMMNS14,BCKMN19,BEP20},
alternative approaches to study ${\mathfrak M}$ and $\chi$ rely on the Nambu-Jona-Lasinio model and extensions thereof
\citep{FS14,PDNS16,FTAPK17}, on the hadron resonance gas (HRG) model
\citep{End13,TDES16,KPB19}, and on yet other techniques
\citep{KLW02,TM14,SC14,KK15,SO15,OS15,TPPYB15,TDH16,AS18a,TDH18,LFL19,KGBHM19,RP19,AFPRT20,AA21}.
Remarkably, a systematic study of the magnetic susceptibility within chiral perturbation (CHPT) -- i.e., the low-energy effective field
theory of QCD -- appears to be lacking.

The overall picture that emerged from these studies is that the QCD vacuum behaves as a diamagnetic medium at low temperatures, while at
higher temperatures, around $T \approx 110 \, \text{MeV}$, it evolves into a paramagnetic medium. The fact that $\chi$ changes from
negative into positive as temperature -- or magnetic field strength -- increase, can be traced back to various reasons. First, at low $T$,
the physics of the system is dominated by the pions that give rise to a negative magnetic susceptibility. However, at more elevated
temperatures, spin-$\frac{1}{2}$ and spin-$1$ hadrons also become important -- unlike the pions they yield positive contributions to
$\chi$. Then, in finite magnetic fields, pions and higher-spin hadrons lead to positive zero-temperature contributions in $\chi$ that grow
as the magnetic field becomes stronger. Overall, as temperature rises, the system undergoes a qualitative change in its particle content:
while hadrons dominate at low temperatures, quarks are the relevant degrees of freedom at high temperatures -- in particular, the
quark-gluon plasma exhibits strong paramagnetic behavior.

Current lattice QCD simulations and most other studies address the temperature regime around or above $T \approx 110 \, \text{MeV}$. An
exception is the hadron resonance gas model: much like chiral perturbation theory it applies at low temperatures. However, the study
in Ref.~\citep{End13} was restricted to noninteracting particles. In general, a quantitative investigation of the diamagnetic and
paramagnetic phases in finite magnetic fields and at low temperatures, is still lacking. In the present two-flavor CHPT analysis, where
pion-pion interactions are taken into account, we provide such a systematic and rigorous low-energy analysis. We derive the two-loop
representation for the magnetization and the magnetic susceptibility at zero and finite temperatures in weak external homogeneous magnetic
fields. We show that the QCD vacuum at $T$=0 is paramagnetic in nonzero magnetic fields. In contrast, the finite-temperature portion in the
magnetic susceptibility is negative, such that the total magnetic susceptibility $\chi_{tot}$ (sum of $T$=0 and finite-$T$ contribution) may
result positive or negative: depending on temperature and magnetic field strength, paramagnetic and diamagnetic phases can be identified in
the low-energy region. In zero magnetic field, $\chi_{tot}$ is strictly negative in the pion-dominated regime, but as the magnetic field
grows, the QCD vacuum turns into a paramagnetic medium at low temperatures. Remarkably, the dependence of $\chi_{tot}$ on temperature is
non-monotonic in the paramagnetic phase. We stress that we provide high-precision results in a parameter domain where lattice QCD
simulations still are a challenge. 

\section{Magnetization}
\label{magnetization}

\begin{figure}
\begin{center}
\includegraphics[width=8.0cm]{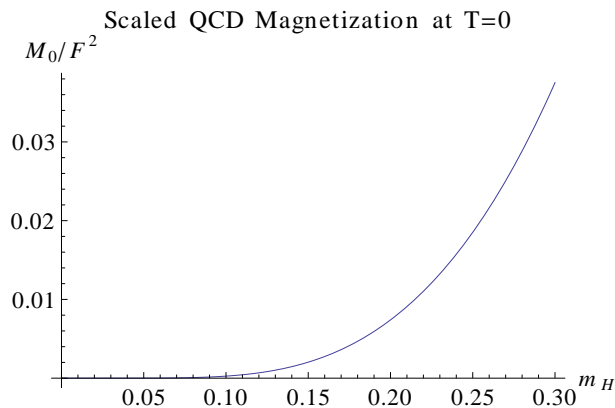}
\end{center}
\caption{Scaled zero-temperature QCD magnetization ${\mathfrak M}_0/F^2$ in terms of magnetic field strength ($m_H$).}
\label{figure1}
\end{figure}

We first consider the magnetization that is induced by the external magnetic field. Based on the representation of the renormalized vacuum
energy density $z^H_0$ given Eq.~(\ref{freeEDp6ZeroTOnlyHtermsBaliScheme})\footnote{For completeness we provide explicit expressions for the
free energy density in Appendix \ref{appendixA}.}, the magnetization at zero temperature amounts to
\begin{eqnarray}
\label{magnetizationp6ZeroT}
{\mathfrak M}_0(M_{\pi},H) & = &
\frac{|qH|}{8 \pi^2} \, {\cal J}_{-2} + \frac{M^2_{\pi}}{16 \pi^2} \, {\cal J}_{-1}
+ \frac{{\overline l}_3}{512 \pi^4} \, \frac{M^4_{\pi}}{F^2} \Big\{ {\cal I}_{-1} + \frac{M^2_{\pi}}{|qH|} \,  {\cal I}_0 \Big\} \nonumber \\
& & + \frac{{\overline l}_3}{1536 \pi^4} \, \frac{M^2_{\pi} |qH|}{F^2} 
- \frac{{\overline l}_6 - {\overline l}_5}{768 \pi^4} \, \frac{|qH|}{F^2} \, \Big\{ 3|qH| {\cal I}_{-1} + M^2_{\pi} {\cal I}_0 \Big\} \, .
\end{eqnarray}
The dimensionless integrals ${\cal I}_n(M_{\pi},H)$ and ${\cal J}_n(M_{\pi},H)$, defined in Eq.~(\ref{dimensionlessIntegralsIJ}), depend on
the pion mass $M_{\pi}$ and the magnetic field $H$. The dimensionless quantities ${\overline l}_3, {\overline l}_5, {\overline l}_6$ are
so-called next-to-leading order low-energy constants. Following Refs.~\citep{GL84,Aok20}, we use the values
${\overline l}_3 = 3.41(82)$ and ${\overline l}_6 - {\overline l}_5 = 2.64 \pm 0.72$. To discuss the properties of the QCD vacuum it is
convenient not to use absolute values of $M_{\pi}, H$ and $T$, but to define dimensionless quantities $m, m_H$, and $t$ as
\begin{equation}
m = \frac{M_{\pi}}{4 \pi F} \, , \qquad m_H = \frac{\sqrt{|q H|}}{4 \pi F} \, , \qquad t = \frac{T}{4 \pi F}\, .
\end{equation}
The denominator $\Lambda_{\chi} \approx 4 \pi F \approx 1 \, \text{GeV}$ is the chiral symmetry breaking scale. For the tree-level pion
decay constant we use $F = 85.6 \, \text{MeV}$ \citep{Aok20}. At low energies, i.e., in the domain where CHPT is valid, the parameters
$m, m_H$, and $t$ are small. In the subsequent plots we choose $t \lessapprox 0.15 \ (T \lessapprox 150 \, \text{MeV})$ and 
$m_H \lessapprox 0.3 \ (|qH| \lessapprox 0.1 \, \text{GeV}^2)$. The dependence of ${\mathfrak M}_0(M_{\pi},H)$ on magnetic field
strength ($m_H$) is illustrated in Fig.~\ref{figure1} for the physically relevant case $M_{\pi} = 140 \, \text{MeV}$ ($m = 0.130$).
${\mathfrak M}_0(M_{\pi},H)$ is positive and grows monotonically as the magnetic field strength increases. The curvature implies
paramagnetic behavior. The limit $H \to 0$ does not pose any problems: $\lim_{H \to 0} {\mathfrak M}_0(M_{\pi},H) = 0$. As expected, no
spontaneous magnetization emerges.

\begin{figure}
\begin{center}
\hbox{
\includegraphics[width=8.0cm]{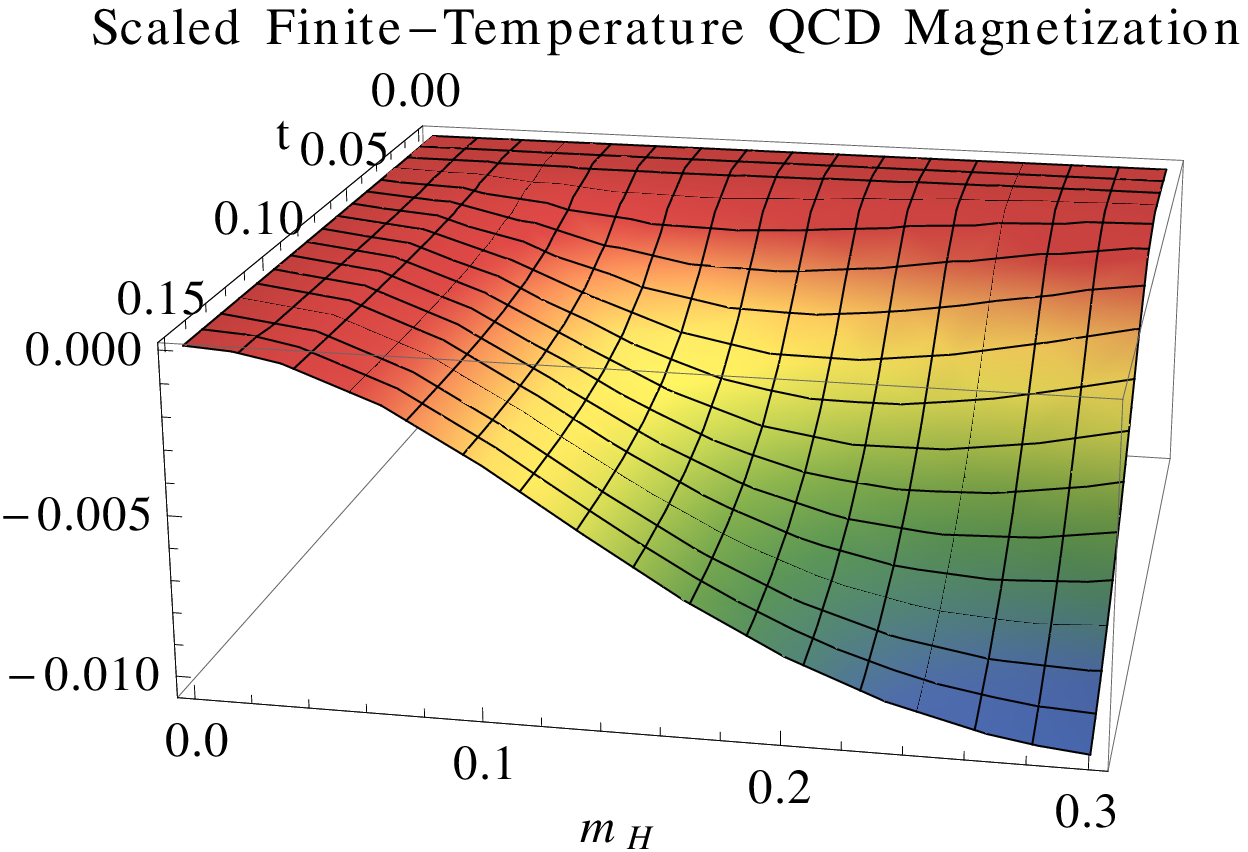}
\includegraphics[width=8.0cm]{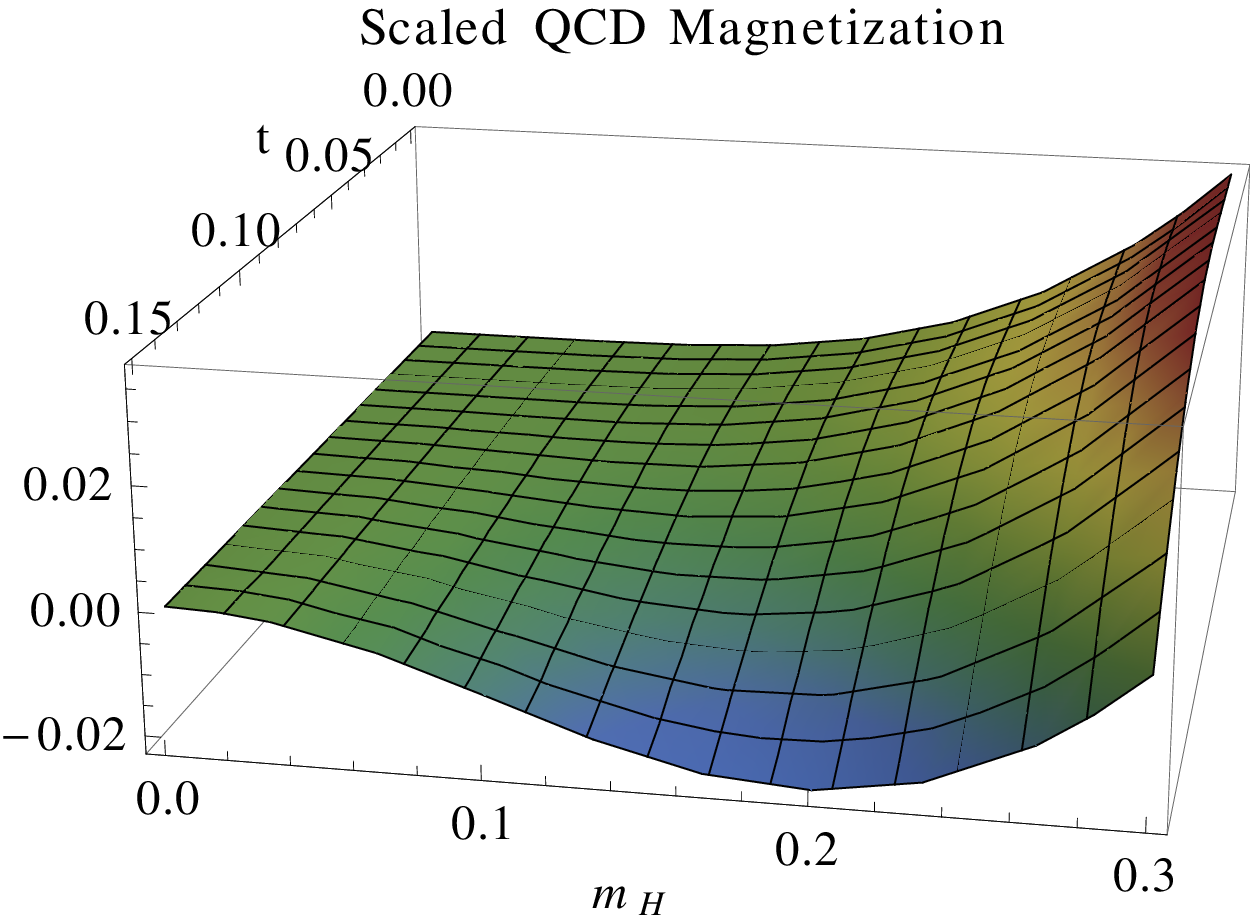}}
\end{center}
\caption{[Color online] Scaled finite-temperature QCD magnetization ${\mathfrak M}_T/T^2$ (LHS) and scaled total QCD magnetization
${\mathfrak M}_{tot}/F^2$ (RHS) in terms of magnetic field strength ($m_H$) and temperature ($t$).}
\label{figure2}
\end{figure}

In contrast to ${\mathfrak M}_0(M_{\pi},H)$, the purely finite-temperature portion ${\mathfrak M}_T(M_{\pi},H)$ in the total
magnetization\footnote{The explicit two-loop representation for ${\mathfrak M}_T(M_{\pi},H)$ can be found in Ref.~\citep{Hof20c}.}
\begin{equation}
{\mathfrak M}_{\text{tot}}(T,M_{\pi},H) = {\mathfrak M}_0(M_{\pi},H) + {\mathfrak M}_T(M_{\pi},H) \, ,
\end{equation}
is negative at $M_{\pi} = 140 \, \text{MeV}$ according to the LHS of Fig.~\ref{figure2}. Its dependence on $T$ and $H$ is nontrivial: at
lower (fixed) temperatures, $|{\mathfrak M}_T(M_{\pi},H)|$ initially grows as the magnetic field gets stronger, goes through a maximum and
then starts to decline. At fixed magnetic field strength, $|{\mathfrak M}_T(M_{\pi},H)|$ also starts to increase less rapidly as temperature
rises. With respect to the (negative) one-loop contribution, the two-loop correction is of the order of a few percent and positive, i.e.,
it slightly weakens the dominant effect. Finally, on the RHS of Fig.~\ref{figure2}, we depict the total magnetization which may take
positive or negative values. Remarkably, this non-monotonic dependence of ${\mathfrak M}_{tot}(T,M_{\pi},H)$ on $H$ implies that the QCD
vacuum may behave as a diamagnetic or paramagnetic medium (see below).

\section{Magnetic Susceptibility}
\label{magneticSusceptibility}

\begin{figure}
\begin{center}
\includegraphics[width=8.0cm]{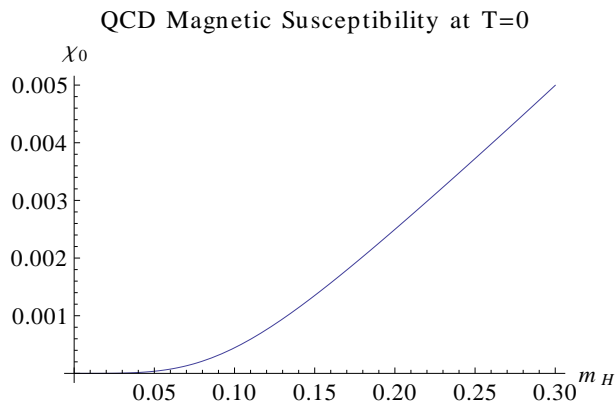}
\end{center}
\caption{Zero-temperature QCD magnetic susceptibility $\chi_0$ in terms of magnetic field strength ($m_H$).}
\label{figure3}
\end{figure}

We now focus on the magnetic susceptibility where we also present zero-temperature and finite-temperature pieces,
\begin{equation}
\chi_{\text{tot}}(T,M_{\pi},H) = \chi_0(M_{\pi},H) + \chi_T(M_{\pi},H) \, ,
\end{equation}
separately. The zero-temperature portion reads
\begin{eqnarray}
\label{magsusp6ZeroT}
{\chi}_0(M_{\pi},H) & = & \frac{1}{16 \pi^2} \Big\{ 2 {\cal J}_{-2} + \frac{2 M^2_{\pi}}{|qH|} \, {\cal J}_{-1} + \frac{M^4_{\pi}}{{|qH|}^2} \,
{\cal J}_0 \Big\}
+ \frac{{\overline l}_3}{512 \pi^4 F^2} \, \frac{M^8_{\pi}}{{|qH|}^3} \, {\cal I}_1 \nonumber \\
& & + \frac{{\overline l}_3}{1536 \pi^4} \, \frac{M^2_{\pi}}{F^2}
- \frac{{\overline l}_6 - {\overline l}_5}{768 \pi^4 F^2} \, \Big\{ 6 |qH| \, {\cal I}_{-1} + 4 M^2_{\pi} \, {\cal I}_0
+ \frac{M^4_{\pi}}{|qH|} \, {\cal I}_1 \Big\} \, .
\end{eqnarray}
According to Appendix \ref{appendixB} where we analyze the limit $|qH| \ll M^2_{\pi}$, the corresponding expansion of ${\chi}_0(M_{\pi},H)$ is
characterized by even powers of the magnetic field,
\begin{equation}
\chi_0(M_{\pi},H) = \alpha_0 + \alpha_2 {|qH|}^2  + \alpha_4 {|qH|}^4 + {\cal O}({|qH|}^6) \, ,
\end{equation}
with coefficients $\alpha_n$ given in Eq.~(\ref{coefficientsAlpha}). As explained in Appendix \ref{appendixA1}, we adopt the standard
renormalization prescription which is to drop in the $T$=0 free energy density all terms quadratic in the magnetic field. This means that
$\chi_0$ in zero magnetic field is set to zero by definition,
\begin{equation}
\lim_{H \to 0} \chi_0(M_{\pi},H) \doteq 0 \qquad \Longleftrightarrow \qquad \alpha_0 \doteq 0 \, .
\end{equation}
The dependence of the zero-temperature magnetic susceptibility on magnetic field strength ($m_H$) at $M_{\pi} = 140 \, \text{MeV}$ is shown
in Fig.~\ref{figure3}: $\chi_0(M_{\pi},H)$ is positive and grows monotonically -- the QCD vacuum at $T$=0 in finite magnetic fields is
paramagnetic.

The finite-temperature portion of the magnetic susceptibility we write as
\begin{equation}
\chi_T(M_{\pi},H) = \chi_1(T,M_{\pi},H) + \chi_2(T,M_{\pi},H) \, .
\end{equation}
The one-loop contribution $\chi_1$ refers to noninteracting pions. The two-loop correction $\chi_2$ contains the pion-pion interaction and
is of the order of a few percent compared to $\chi_1$. The respective expressions are rather lengthy and provided in Appendix
\ref{appendixC}. On the LHS of Figure \ref{figure4} we depict the dependence of $\chi_T(M_{\pi},H)$ on temperature and magnetic field at
$M_{\pi} = 140 \, \text{MeV}$: it is negative in most of parameter space accessible by CHPT -- only in stronger magnetic fields it takes
slightly positive values. At fixed $T$, $\chi_T(M_{\pi},H)$ increases as the magnetic field becomes stronger and then reaches a plateau. At
fixed $H$, overall, $\chi_T(M_{\pi},H)$ decreases as temperature rises -- however, in stronger magnetic fields, it first slightly grows and
then falls off to negative values -- hence exhibiting non-monotonic behavior.

\begin{figure}
\begin{center}
\hbox{
\includegraphics[width=8.0cm]{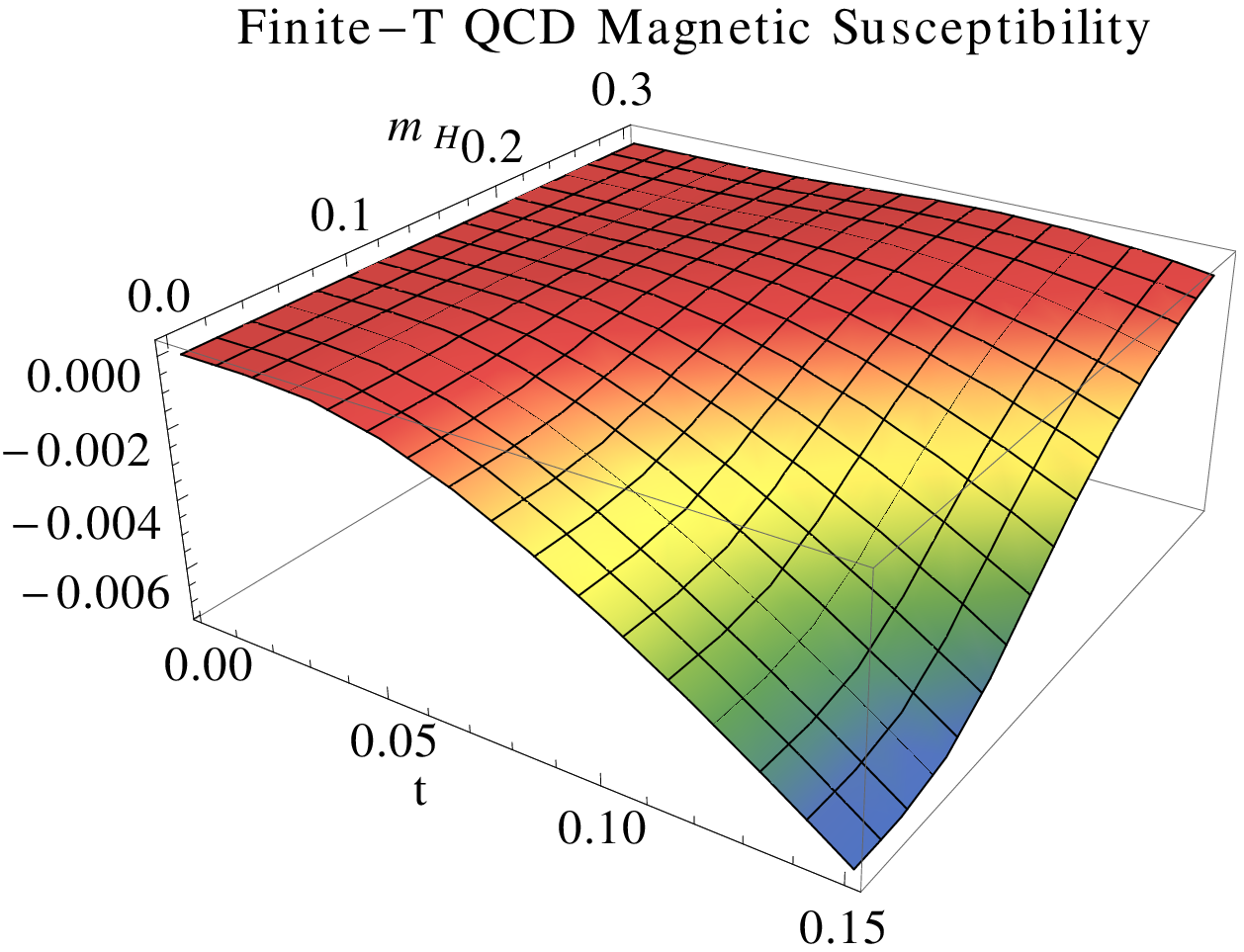}
\includegraphics[width=8.0cm]{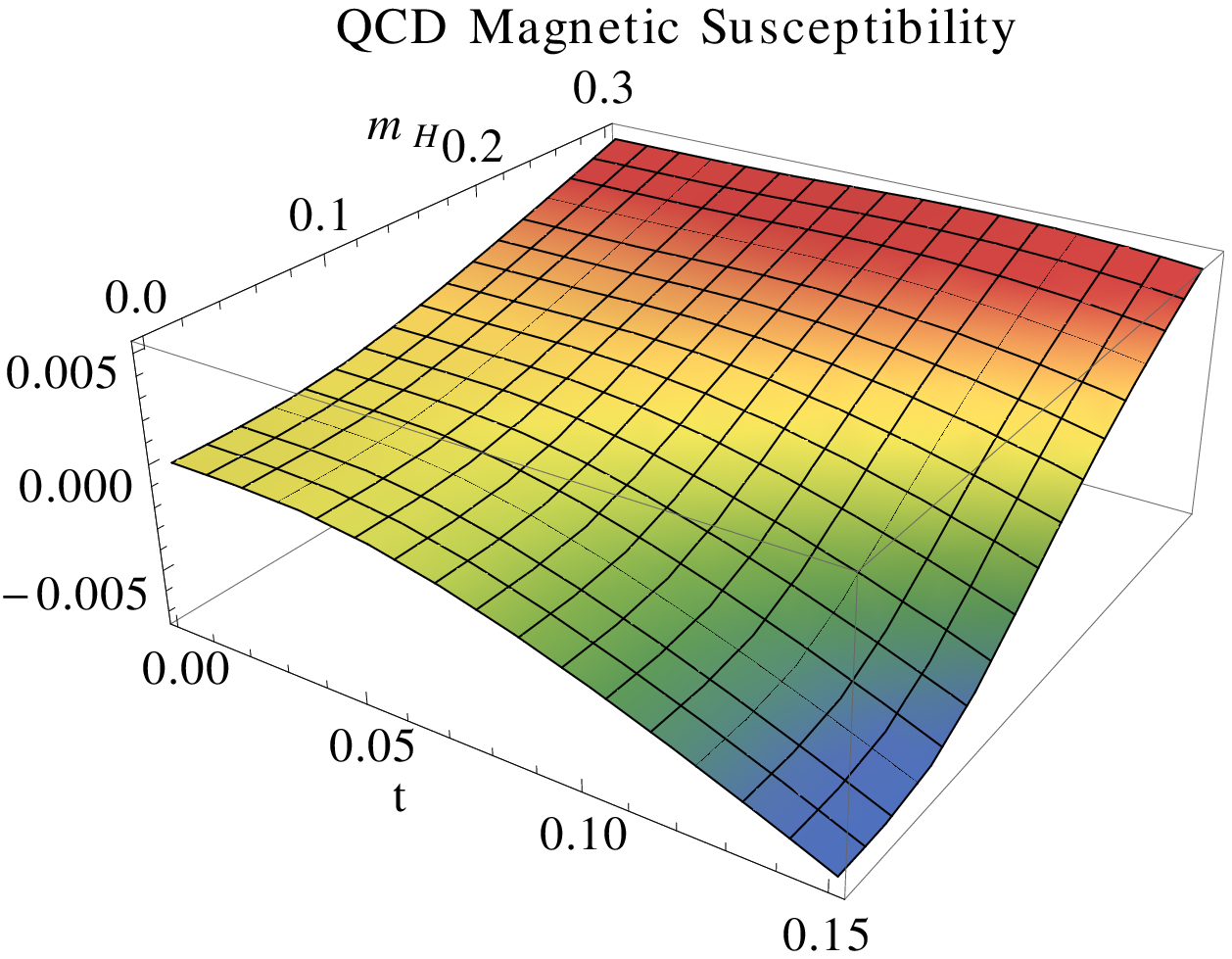}}
\end{center}
\caption{[Color online] Finite-temperature QCD magnetic susceptibility $\chi_T$ (LHS) and total QCD magnetic susceptibility $\chi_{tot}$
(RHS) in terms of magnetic field strength ($m_H$) and temperature ($t$).}
\label{figure4}
\end{figure}

\begin{figure}
\begin{center}
\includegraphics[width=8.0cm]{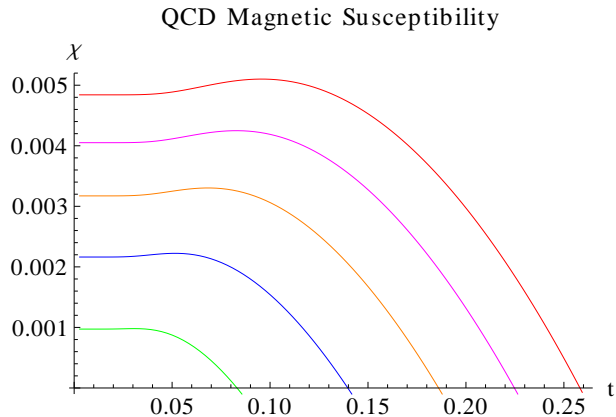}
\end{center}
\caption{[Color online] Non-monotonic behavior of the total QCD magnetic susceptibility in the paramagnetic phase: Dependence of
$\chi_{tot}$ on temperature ($t$) at fixed magnetic field strength $|qH|=\{ 0.02,0.04,0.06,0.08,0.1 \} \, {\text{GeV}}^2$ (bottom to top).}
\label{figure5}
\end{figure}

The main result of the present study concerns the total magnetic susceptibility $\chi_{tot}(T,M_{\pi},H)$, i.e., the superposition of
$\chi_0(M_{\pi},H)$ and $\chi_T(M_{\pi},H)$. This quantity indeed exhibits some remarkable features in the low-energy region. First, as we
illustrate on the RHS of Figure \ref{figure4}, in stronger magnetic fields, the QCD vacuum -- irrespective of temperature -- is
{\it paramagnetic}. In weaker magnetic fields, however, a {\it diamagnetic} phase starts to emerge -- eventually, at $H$=0, the QCD vacuum
is diamagnetic in the entire regime $t < 0.15$. Second, the behavior of $\chi_{tot}(T,M_{\pi},H)$ in the paramagnetic phase is non-monotonic
as can be better appreciated in Figure \ref{figure5}: as temperature grows -- while $H$ kept fixed -- $\chi_{tot}$ initially rises, goes
through a maximum and then starts to drop. It should be pointed out that this phenomenon already emerges at one-loop order and is slightly
weakened (of the order of a few permille) by the two-loop correction. The effect is more pronounced in stronger magnetic fields and it is
absent at $H$=0 -- in the latter case the QCD vacuum is purely diamagnetic in the pion-dominated low-temperature phase. Third, the behavior
of $\chi_{tot}(T,M_{\pi},H)$ is also non-monotonic in the diamagnetic phase. Extrapolating our results to higher temperatures, $\chi_{tot}$
eventually develops a minimum and then starts to rise as $T$ increases -- this latter effect is not shown in Figure \ref{figure4} since we
are about to leave the low-energy domain where CHPT is valid. The point is that the two-loop correction becomes large and positive at
higher $T$ and weak magnetic fields. While the (negative) one-loop contribution at $H$=0 simply implies a monotonic decrease of $\chi_{tot}$
with $T$, the two-loop contribution causes the non-monotonic behavior, signaling that the diamagnetic QCD vacuum eventually turns into
paramagnetic at higher temperatures.

\section{Conclusions}
\label{conclusions}

The subtle interplay between zero-$T$ and finite-$T$ contributions leads to the nontrivial behavior of the magnetization and magnetic
susceptibility that we observe at low temperatures and weak magnetic fields. The comparison of CHPT studies of the quark condensate with
lattice data performed in Ref.~\citep{BBEFKKSS12b} suggests that CHPT is perfectly valid up to magnetic field strengths of
$|qH| \lessapprox 0.1 \, \text{GeV}^2 \, (m_H  \lessapprox 0.3)$. Likewise, the HRG model analysis of Ref.~\citep{End13} concludes that
pions no longer dominate at low temperatures beyond $|qH| \gtrapprox 0.2 \, \text{GeV}^2 \, (m_H  \gtrapprox 0.4)$. Our high-precision and
fully systematic results for the magnetization and magnetic susceptibility are within this parameter range and thus accurately describe the
pion-dominated phase. We hence complement and extend all previous studies on these two observables to a parameter domain that is hardly
accessible by lattice QCD at present and has not been examined by any other method beyond leading order in a systematic way. It remains to
be seen whether future lattice QCD simulations can quantitatively explore the diamagnetic and paramagnetic phases in the low-energy region
of QCD and confirm our predictions.

\section*{Acknowledgments}

The author thanks G.\ S.\ Bali, J.\ Bijnens, G.\ Endr\"odi, and H.\ Leutwyler for correspondence.

\begin{appendix}

\section{Representation of the Free Energy Density}
\label{appendixA}

The purpose of this Appendix is to make the presentation self-contained by providing explicit expressions for the two-loop free energy
density $z$ which is the starting point of our analysis. It is convenient to split $z$ into two pieces,
\begin{equation}
z = z_0 + z_T \, ,
\end{equation}
where $z_0$ is the free energy density at $T$=0 and $z_T$ is the finite-temperature portion. In what follows we discuss these two pieces
individually. The two-loop calculation within the framework of two-flavor chiral perturbation theory\footnote{Outlines of chiral
perturbation theory are given in Refs.~\citep{Leu95,Sch03}.} in the isospin limit $m_u = m_d$ was performed in Refs.~\citep{Hof20a,Hof20b}.

\subsection{Zero Temperature}
\label{appendixA1}

The renormalized vacuum energy density $z_0$ takes the form
\begin{eqnarray}
\label{freeEDp6ZeroT}
z_0 & = & - F^2 M^2 + \frac{M^4}{64 \pi^2} \, \Big( {\overline l_3} - 4{\overline h_1} - \frac{3}{2} \Big) + \frac{{|qH|}^2}{96 \pi^2} \,
( {\overline h_2} - 1) - \frac{{|qH|}^2}{16 \pi^2}  {\cal J}_{-2} \nonumber \\
& &  + \frac{3{\overline l}_3 ({\overline c}_{10} + 2 {\overline c}_{11})}{1024 \pi^4} \, \frac{M^6}{F^2}
- \frac{({\overline l}_6 - {\overline l}_5){\overline c}_{34}}{768 \pi^4} \, \frac{{|qH|}^2 M^2}{F^2} \nonumber \\
& & - \frac{{\overline l}_3}{512 \pi^4} \,  \frac{M^4 |qH|}{F^2} \, {\cal I}_{-1}
+ \frac{{\overline l}_6 - {\overline l}_5}{768 \pi^4} \, \frac{{|qH|}^3}{F^2} \, {\cal I}_{-1} + {\cal O}(p^8) \, ,
\end{eqnarray}
where the integrals ${\cal I}_n$ and ${\cal J}_n$ are
\begin{eqnarray}
\label{dimensionlessIntegralsIJ}
{\cal I}_n & = & {\int}_{\!\!\! 0}^{\infty} \mbox{d} \rho \, \rho^n \, \exp\Big( -\frac{M^2}{|qH|} \rho \Big) \,
\Big( \frac{1}{\sinh(\rho)} - \frac{1}{\rho} \Big) \, , \nonumber \\
{\cal J}_n & = & {\int}_{\!\!\! 0}^{\infty} \mbox{d} \rho \, \rho^n \, \exp\Big( -\frac{M^2}{|qH|} \rho \Big) \,
\Big( \frac{1}{\sinh(\rho)} - \frac{1}{\rho} + \frac{\rho}{6} \Big) \, .
\end{eqnarray}
$z_0$ contains the renormalized next-to-leading order and next-to-next-to-leading order effective constants,
${\overline l}_3, {\overline l}_5, {\overline l}_6,{\overline h}_1, {\overline h}_2$ and
${\overline c}_{10}, {\overline c}_{11}, {\overline c}_{34}$, respectively. Details on the definition and running of these low-energy
couplings can be found in Appendix A of Ref.~\citep{Hof20b}, as well as in the original references \citep{GL84,BCE00}. Numerical values are
provided in the main body of the present paper. Finally, $M$ is the tree-level pion mass which is related to the (one-loop) pion mass
$M_{\pi}$ as
\begin{equation}
\label{Mpi}
M^2_{\pi} = M^2 - \frac{{\overline l}_3}{32 \pi^2} \, \frac{M^4}{F^2} + {\cal O}(M^6) \, .
\end{equation}

A crucial question is which contributions in the $T$=0 free energy density are physically relevant. Since we are interested in how the QCD
vacuum is affected by the magnetic field, we can ignore all terms that do not involve the magnetic field. We are then left with
\begin{eqnarray}
\label{freeEDp6ZeroTOnlyHterms}
{\tilde z}^H_0 & = & \frac{{|qH|}^2}{96 \pi^2} \, ( {\overline h_2} - 1)
-  \frac{{|qH|}^2}{16 \pi^2}  {\cal J}_{-2}
- \frac{({\overline l}_6 - {\overline l}_5){\overline c}_{34}}{768 \pi^4} \, \frac{{|qH|}^2 M^2}{F^2} \nonumber \\
& & - \frac{{\overline l}_3}{512 \pi^4} \,  \frac{M^4 |qH|}{F^2} \, {\cal I}_{-1}
+ \frac{{\overline l}_6 - {\overline l}_5}{768 \pi^4} \, \frac{{|qH|}^3}{F^2} \, {\cal I}_{-1} \, .
\end{eqnarray}
Clearly, the contribution
\begin{equation}
\label{drop0}
-\frac{{|qH|}^2}{96 \pi^2}
\end{equation}
can be dropped as it is independent of the properties of the pions: it does not involve the pion mass, but solely depends on the external
magnetic field. Note that there are further terms quadratic in the magnetic field. At chiral order $p^4$, we have
\begin{equation}
\label{drop1}
\frac{{\overline h}_2}{96 \pi^2} \, {|qH|}^2 \, ,
\end{equation}
and at chiral order $p^6$ we have
\begin{equation}
\label{drop2}
- \frac{({\overline l}_6 - {\overline l}_5){\overline c}_{34}}{768 \pi^4} \, \frac{M^2}{F^2} \, {|qH|}^2\, .
\end{equation}
Then -- according to the analysis of the integral ${\cal I}_{-1}$ in the limit $|qH| \ll M^2$ performed in Appendix \ref{appendixB} -- an
additional term quadratic in the magnetic field arises at chiral order $p^6$, namely,
\begin{equation}
\label{drop3}
\frac{{\overline l}_3}{3072 \pi^4} \, \frac{M^2}{F^2} \, {|qH|}^2 \, .
\end{equation}
In order to compare our results with the literature, we adopt the renormalization prescription for the zero-temperature free energy density
that underlies lattice as well HRG model studies
\citep{BBCCEKPS12c,BEMNS13,BBEGS13,BBES13,BEMNS13b,BBEKS14,BEMNS14,BBES14,LT14,End14,BEMMNS14,BCKMN19,BEP20,End13,SC14,TDES16,KPB19},
which is to {\bf drop in the $T$=0 free energy density all terms quadratic in the magnetic field}. In our CHPT framework this corresponds
to subtracting the terms (\ref{drop0})-(\ref{drop3}) from the vacuum energy density Eq.~(\ref{freeEDp6ZeroTOnlyHterms}). The properly
normalized zero-temperature free energy density hence takes the form
\begin{eqnarray}
\label{freeEDp6ZeroTOnlyHtermsBaliScheme}
z^H_0 & = & - \frac{{|qH|}^2}{16 \pi^2}  {\cal J}_{-2}
- \frac{{\overline l}_3}{512 \pi^4} \,  \frac{M^4 |qH|}{F^2} \, \Big( {\cal I}_{-1} + \frac{|qH|}{6 M^2} \Big)
+ \frac{{\overline l}_6 - {\overline l}_5}{768 \pi^4} \, \frac{{|qH|}^3}{F^2} \, {\cal I}_{-1} \, .
\end{eqnarray}
Note that the difference between the tree-level pion mass $M$ and the one-loop pion mass $M_{\pi}$ only starts manifesting itself beyond
chiral order $p^6$. We can therefore safely replace $M$ by $M_{\pi}$ in Eq.~(\ref{freeEDp6ZeroTOnlyHtermsBaliScheme}). The expansion of
$z^H_0$ in the limit $|qH| \ll M^2$ gives rise to even powers of the magnetic field that start at order ${|qH|}^4$ -- all terms quadratic
in the magnetic field have been eliminated.\footnote{This means that the NLO effective constant ${\overline h}_2$ and the NNLO effective
constant ${\overline c}_{34}$ are irrelevant within the adopted renormalization prescription.} Equivalently, within this convention, the
$T$=0 magnetic susceptibility $\chi_0$ in zero magnetic field is set to zero by definition,
\begin{equation}
\lim_{H \to 0} \chi_0 = \lim_{H \to 0} - {\frac{\mbox{d}^2 z^H_0}{\mbox{d} {|qH|}^2}} \doteq 0 \, .
\end{equation}
It is important to emphasize that our assessment of whether the QCD vacuum has diamagnetic or paramagnetic properties is tied to this
renormalization convention.

\subsection{Finite Temperature}
\label{appendixA2}

For completeness we provide the finite-temperature contribution $z_T$ in the free energy density. Following Ref.~\citep{Hof20a}, the
two-loop representation reads
\begin{eqnarray}
\label{fedPhysicalM}
z_T & = & - g_0(M^{\pm}_{\pi},T,0) -\mbox{$ \frac{1}{2}$} g_0(M^0_{\pi},T,0)- {\tilde g}_0(M^{\pm}_{\pi},T,H) \nonumber \\
& & + \frac{M^2_{\pi}}{2 F^2} \, g_1(M^{\pm}_{\pi},T,0) \, g_1(M^0_{\pi},T,0)
- \frac{M^2_{\pi}}{8 F^2} \, {\Big\{ g_1(M^0_{\pi},T,0)  \Big\}}^2 \nonumber \\
& & + \frac{M^2_{\pi}}{2 F^2} \, g_1(M^0_{\pi},T,0) \, {\tilde g}_1(M^{\pm}_{\pi},T,H) + {\cal O}(p^8) \, ,
\end{eqnarray}
with respective Bose functions defined as
\begin{eqnarray}
\label{BoseFunctions}
g_r({\cal M},T,0) & = & \frac{T^{4-2r}}{{(4 \pi)}^r} \, {\int}_{\!\!\! 0}^{\infty}  \mbox{d} \rho \, \rho^{r-3} \,
\exp\Big( -\frac{{\cal M}^2}{4 \pi T^2} \rho \Big) \Bigg[ S\Big( \frac{1}{\rho} \Big) -1 \Bigg] \, , \nonumber \\
{\tilde g}_r(M^{\pm}_{\pi},T,H) & = & \frac{T^{2-2r}}{{(4 \pi)}^{r+1}} \, |qH| {\int}_{\!\!\! 0}^{\infty} \mbox{d} \rho \, \rho^{r-2} \,
\Bigg( \frac{1}{\sinh(|qH| \rho /4 \pi T^2)} - \frac{4 \pi T^2}{|qH| \rho} \Bigg) \nonumber \\
& & \times \, \exp\Big( -\frac{{(M^{\pm}_{\pi})}^2}{4 \pi T^2} \rho \Big) \Bigg[ S\Big( \frac{1}{\rho} \Big) -1 \Bigg] \, ,
\end{eqnarray}
where $S(z)$,
\begin{equation}
S(z) = \sum_{n=-\infty}^{\infty} \exp(- \pi n^2 z) \, ,
\end{equation}
is the Jacobi theta function. The Bose functions involve the quantities $M^{\pm}_{\pi}$ and $M^0_{\pi}$, i.e., the masses of the charged and
neutral pions subjected to the magnetic field,
\begin{eqnarray}
\label{chargedNeutralPionMass}
{(M^{\pm}_{\pi})}^2 & = & M^2_{\pi} + \frac{{\overline l}_6 - {\overline l}_5}{48 \pi^2} \, \frac{{|qH|}^2}{F^2} \, , \nonumber \\
{(M^0_{\pi})}^2 & = & M^2_{\pi}  + \frac{M^2 |qH|}{16 \pi^2 F^2} \, {\cal I}_{-1} \, .
\end{eqnarray}
The symbol $\cal M$ in the kinematical Bose functions $g_r$ can either denote $M^{\pm}_{\pi}$ or $M^0_{\pi}$ depending on context. The
quantity $M_{\pi}$ is the (one-loop) pion mass in zero magnetic field defined by Eq.~(\ref{Mpi}).

\section{Zero-Temperature Magnetic Susceptibility in the Limit $|qH| \ll M^2$}
\label{appendixB}

In this Appendix we identify the structure of magnetic field powers in the zero-temperature magnetic susceptibility ${\chi}_0(M,H)$ by
considering the limit $|qH| \ll M^2$. The quantity ${\chi}_0(M,H)$,
\begin{equation}
\chi_0(M,H) = - {\frac{\mbox{d}^2 z_0}{\mbox{d} {|qH|}^2}} \, ,
\end{equation}
with $z_0$ given in Eq.~(\ref{freeEDp6ZeroT}), contains the integrals ${\cal I}_n$ and ${\cal J}_n$ -- see
Eq.~(\ref{dimensionlessIntegralsIJ}) -- that we write as
\begin{eqnarray}
{\cal I}_n & = & \varepsilon^{n+1} \, {\int}_{\!\!\! 0}^{\infty} \mbox{d} \rho \, \rho^n \, e^{- \rho} \,
\Big( \frac{1}{\sinh(\varepsilon \rho)} - \frac{1}{\varepsilon \rho} \Big) \, , \nonumber \\
{\cal J}_n & = & \varepsilon^{n+1} \, {\int}_{\!\!\! 0}^{\infty} \mbox{d} \rho \, \rho^n \, e^{- \rho} \,
\Big( \frac{1}{\sinh(\varepsilon \rho)} - \frac{1}{\varepsilon \rho} + \frac{\varepsilon \rho}{6} \Big) \, ,
\end{eqnarray}
where $\varepsilon$,
\begin{equation}
\varepsilon = \frac{|qH|}{M^2} \, ,
\end{equation}
is the relevant expansion parameter. The integrands yield the series
\begin{equation}
\frac{1}{\sinh(\varepsilon \rho)} - \frac{1}{\varepsilon \rho} = {\hat c}_1 \rho \, \varepsilon + {\hat c}_2 \rho^3 \varepsilon^3
+ {\hat c}_3 \rho^5 \varepsilon^5 + {\cal O}(\varepsilon^7) \, ,
\end{equation}
with the first five Taylor coefficients as
\begin{eqnarray}
& {\hat c}_1 & = - \frac{1}{6} \approx -0.167 \, , \nonumber \\
& {\hat c}_2 & = \frac{7}{360 } \approx 0.0194 \, , \nonumber \\
& {\hat c}_3 & = - \frac{31}{15 120} \approx -0.00205 \, , \nonumber \\
& {\hat c}_4 & = \frac{127}{604 800} \approx 0.000210 \, , \nonumber \\
& {\hat c}_5 & = -\frac{73}{3 421 440} \approx -0.0000213 \, . 
\end{eqnarray}
Collecting terms, we find that the zero-temperature magnetic susceptibility features even powers of the magnetic field and amounts to
\begin{equation}
\chi_0(M,H) = \alpha_0 + \alpha_2 {|qH|}^2  + \alpha_4 {|qH|}^4 + {\cal O}({|qH|}^6) \, ,
\end{equation}
with respective coefficients
\begin{eqnarray}
\label{coefficientsAlpha}
\alpha_0 & = & - \frac{{\overline h}_2 -1}{48 \pi^2} + \frac{{\overline c}_{34} ({\overline l}_6 - {\overline l}_5)}{384 \pi^4} \,
\frac{M^2}{F^2} - \frac{{\overline l}_3}{1536 \pi^4} \, \frac{M^2}{F^2} \, , \nonumber \\
\alpha_2 & = & \frac{7}{480 \pi^2 M^4} + \frac{7 {\overline l}_3}{7680 \pi^4 F^2 M^2}
+ \frac{{\overline l}_6 - {\overline l}_5}{384 \pi^4 F^2 M^2} \, , \nonumber \\
\alpha_4 & = & - \frac{31}{1344 \pi^2 M^8} - \frac{31 {\overline l}_3}{10752 \pi^4 F^2 M^6}
- \frac{7 ({\overline l}_6 - {\overline l}_5)}{4608 \pi^4 F^2 M^6} \, .
\end{eqnarray}
Again we point out that the renormalization convention adapted for the $T$=0 free energy density specified in Appendix \ref{appendixA1}
implies that the coefficient $\alpha_0$ is set to zero by definition, i.e., the contribution $\alpha_0$ is subtracted from the
zero-temperature magnetic susceptibility $\chi_0(M,H)$.

\section{Magnetic Susceptibility at Finite Temperature}
\label{appendixC}

Here we provide the explicit representation of the finite-temperature magnetic susceptibility -- $\chi_T$ -- in terms of the various
kinematical Bose functions involved.\footnote{The Bose functions $g_r$ and ${\tilde g}_r$ are defined in Eq.~(\ref{BoseFunctions}).} In the
derivation of $\chi_T$,
\begin{equation}
\chi_T = -\frac{{\mbox{d}}^2 z_T}{\mbox{d} {|qH|}^2} \, ,
\end{equation}
the following identities featuring derivatives of Bose functions with respect to the magnetic field are useful,
\begin{eqnarray}
\frac{\mbox{d}}{\mbox{d} |qH|} \, g_r(M^{\pm}_{\pi},T,0) & = & -\frac{{\overline l}_6 - {\overline l}_5}{24 \pi^2} \, \frac{|qH|}{F^2} \,
g_{r+1}(M^{\pm}_{\pi},T,0) \, , \nonumber \\
\frac{\mbox{d}}{\mbox{d} |qH|} \, g_r(M^0_{\pi},T,0) & = & -\frac{M^2_{\pi}}{16 \pi^2 F^2} \,
\Big( {\cal I}_{-1} + \frac{M^2_{\pi}}{|qH|} \, {\cal I}_0  \Big) \, g_{r+1}(M^0_{\pi},T,0) \, , \nonumber \\
\frac{\mbox{d}}{\mbox{d} |qH|} \, {\tilde g}_r(M^{\pm}_{\pi},T,H) & = & \frac{1}{|qH|} \, {\tilde g}_r(M^{\pm}_{\pi},T,H)
- \frac{{\overline l}_6 -{\overline l}_5}{24 \pi^2} \, \frac{|qH|}{F^2} \, {\tilde g}_{r+1}(M^{\pm}_{\pi},T,H) \nonumber \\
& & + {\tilde g}^{[H]}_r(M^{\pm}_{\pi},T,H) \, , 
\end{eqnarray}
where
\begin{eqnarray}
{\tilde g}^{[H]}_r(M^{\pm}_{\pi},T,H) & = & \frac{T^{2-2r}}{{(4 \pi)}^{r+1}} \, |qH| {\int}_{\!\!\! 0}^{\infty} \mbox{d} \rho \rho^{r-2} \,
\Bigg( -\frac{\rho \coth(|qH| \rho /4 \pi T^2)}{4 \pi T^2 \sinh(|qH| \rho /4 \pi T^2)} + \frac{4 \pi T^2}{{|qH|}^2 \rho} \Bigg) \nonumber \\
& & \times \, \exp\Big( -\frac{{(M^{\pm}_{\pi})}^2}{4 \pi T^2} \rho \Big) \Bigg[ S\Big( \frac{1}{\rho} \Big) -1 \Bigg] \, .
\end{eqnarray}
Instead of using dimensionful Bose functions $g_r$ and ${\tilde g}_r$, it is more transparent to express observables by dimensionless Bose
functions $h_r$ and ${\tilde h}_r$, 
\begin{equation}
\label{conversion}
h_0 = \frac{g_0}{T^4} \, , \quad  {\tilde h}_0 = \frac{{\tilde g}_0}{T^4} \, , \qquad
h_1 = \frac{g_1}{T^2} \, , \quad  {\tilde h}_1 = \frac{{\tilde g}_1}{T^2} \, , \qquad
h_2 = g_2 \, , \quad  {\tilde h}_2 = {\tilde g}_2 \, .
\end{equation}
The finite-temperature magnetic susceptibility $\chi_T$ can then be written in the form
\begin{equation}
\chi_T(M_{\pi},H) = \chi_1(T,M_{\pi},H) + \chi_2(T,M_{\pi},H) \, .
\end{equation}
One-loop and two-loop contributions, respectively, are
\begin{eqnarray}
\label{magsusT}
\chi_1 & = & \frac{m_{\pm} t^2}{m_H^2} \, h_1(M^{\pm}_{\pi},T,0) + m_{\pm}^2 \, h_2(M^{\pm}_{\pi},T,0)
+ \frac{t^2}{4 m_H} \, \chi_0 \, h_1(M^0_{\pi},T,0) \nonumber \\
& & + \frac{m_0^2}{2} \, h_2(M^0_{\pi},T,0) + \frac{3 m_{\pm} t^2}{m_H^2} \, {\tilde h}_1(M^{\pm}_{\pi},T,H)
+ \frac{t^2}{m_H^2} \, {\tilde h}^{[H]}_0(M^{\pm}_{\pi},T,H) \nonumber \\
& & + \frac{t^2}{2 m_H} \, \frac{\mbox{d}}{\mbox{d} m_H} \, {\tilde h}^{[H]}_0(M^{\pm}_{\pi},T,H)
+ m_{\pm}^2 {\tilde h}_2(M^{\pm}_{\pi},T,H) + m_{\pm} {\tilde h}^{[H]}_1(M^{\pm}_{\pi},T,H) \, , \nonumber \\
\chi_2 & = & 8 \pi^2 m^2 \Bigg\{ - \frac{t^2}{m^2_H} \, m_{\pm} h_2(M^{\pm}_{\pi},T,0) h_1(M^0_{\pi},T,0)
- m_{\pm}^2 h_3(M^{\pm}_{\pi},T,0) h_1(M^0_{\pi},T,0) \nonumber \\
& & - 2 m_{\pm} m_0 h_2(M^{\pm}_{\pi},T,0) h_2(M^0_{\pi},T,0)
- \frac{t^2}{2 m_H} \, \chi_0 \, h_1(M^{\pm}_{\pi},T,0) h_2(M^0_{\pi},T,0) \nonumber \\
& & - m_0^2 h_1(M^{\pm}_{\pi},T,0) h_3(M^0_{\pi},T,0)
- \frac{t^2}{2 m_H} \, \chi_0 \, h_2(M^0_{\pi},T,0) {\tilde h}_1(M^{\pm}_{\pi},T,H) \nonumber \\
& & - m_0^2 h_3(M^0_{\pi},T,0) {\tilde h}_1(M^{\pm}_{\pi},T,H)
- \frac{2 t^2}{m^2_H} \, m_0 h_2(M^0_{\pi},T,0) {\tilde h}_1(M^{\pm}_{\pi},T,H) \nonumber \\
& & - m_0 m_{\pm} h_2(M^0_{\pi},T,0) {\tilde h}_2(M^{\pm}_{\pi},T,H)
- 2 m_0 h_2(M^0_{\pi},T,0) {\tilde h}^{[H]}_1(M^{\pm}_{\pi},T,H) \nonumber \\
& & - \frac{2 t^2}{m^2_H} \,  m_{\pm} \, h_1(M^0_{\pi},T,0) {\tilde h}_2(M^{\pm}_{\pi},T,H)
- \frac{t^2}{m^2_H} \, h_1(M^0_{\pi},T,0) {\tilde h}^{[H]}_1(M^{\pm}_{\pi},T,H) \nonumber \\
& & - m_{\pm}^2 h_1(M^0_{\pi},T,0) {\tilde h}_3(M^{\pm}_{\pi},T,H)
- m_{\pm} h_1(M^0_{\pi},T,0) {\tilde h}^{[H]}_2(M^{\pm}_{\pi},T,H) \nonumber \\
& & - \frac{t^2}{2 m_H} \, h_1(M^0_{\pi},T,0) \frac{\mbox{d}}{\mbox{d} m_H} \, {\tilde h}^{[H]}_1(M^{\pm}_{\pi},T,H)
+ \frac{m_0^2}{2} \, h_3(M^0_{\pi},T,0) h_1(M^0_{\pi},T,0) \nonumber \\
& & + \frac{t^2}{4 m_H} \, \chi_0\, h_2(M^0_{\pi},T,0) h_1(M^0_{\pi},T,0)
+ \frac{m_0^2}{2} \, h_2(M^0_{\pi},T,0) h_2(M^0_{\pi},T,0) \Bigg\} \, ,
\end{eqnarray}
with coefficients
\begin{eqnarray}
m_{\pm} & = & - \frac{2({\overline l}_6 - {\overline l}_5) m^2_H}{3} \, , \nonumber \\
m_0 & = & - m^2 {\int}_{\!\!\! 0}^{\infty} \mbox{d} \rho \, \rho^{-1} \, \exp\Big( -\frac{m^2}{m^2_H} \rho \Big) \,
\Big( \frac{1}{\sinh(\rho)} - \frac{1}{\rho} \Big) \nonumber \\
& & - \frac{m^4}{m^2_H} \, {\int}_{\!\!\! 0}^{\infty} \mbox{d} \rho \,\exp\Big( -\frac{m^2}{m^2_H} \rho \Big) \,
\Big( \frac{1}{\sinh(\rho)} - \frac{1}{\rho} \Big) \, , \nonumber \\
\chi_0 & = & - \frac{2 m^6}{m^5_H} \, {\int}_{\!\!\! 0}^{\infty} \mbox{d} \rho \, \rho \, \exp\Big( -\frac{m^2}{m^2_H} \rho \Big) \,
\Big( \frac{1}{\sinh(\rho)} - \frac{1}{\rho} \Big) \, .
\end{eqnarray}
The dimensionless functions ${\tilde h}^{[H]}_r(M^{\pm}_{\pi},T,H)$ read
\begin{eqnarray}
& & {\tilde h}^{[H]}_0(M^{\pm}_{\pi},T,H) = \frac{{\tilde g}^{[H]}_0(M^{\pm}_{\pi},T,H)}{T^2} \, , \qquad
{\tilde h}^{[H]}_1(M^{\pm}_{\pi},T,H) = {\tilde g}^{[H]}_1(M^{\pm}_{\pi},T,H) \, , \nonumber \\
& & {\tilde h}^{[H]}_2(M^{\pm}_{\pi},T,H) = {\tilde g}^{[H]}_2(M^{\pm}_{\pi},T,H) \, \times T^2\, ,
\end{eqnarray}
while $h_3$ and ${\tilde h}_3$ are
\begin{equation}
\label{conversion2}
h_3 = g_3 \, T^2 \, , \quad  {\tilde h}_3 = {\tilde g}_3 \, T^2 \, .
\end{equation}

\end{appendix}

\end{document}